\documentclass[aps,prl,twocolumn,showpacs,superscriptaddress,amsmath,longbibliography,amssymb]{revtex4-1}
\usepackage{graphicx}
\usepackage{color,soul}
\usepackage{xcolor}

\begin{document}

\title{Optimal Number of Faces for Fast Self-Folding Kirigami}

\author{H. P. M. Melo} \email{hpmelo@fc.ul.pt}
\affiliation{Centro de F\'{i}sica Te\'{o}rica e Computacional, Universidade de Lisboa, 1749-016
Lisboa, Portugal}

\author{C. S. Dias} \email{csdias@fc.ul.pt}
\affiliation{Centro de F\'{i}sica Te\'{o}rica e Computacional, Universidade de Lisboa, 1749-016
Lisboa, Portugal} 
\affiliation{Departamento de F\'{\i}sica, Faculdade de Ci\^{e}ncias,
Universidade de Lisboa, 1749-016 Lisboa, Portugal} 

\author{N. A. M. Ara\'ujo} \email{nmaraujo@fc.ul.pt}
\affiliation{Centro de F\'{i}sica Te\'{o}rica e Computacional, Universidade de Lisboa, 1749-016
Lisboa, Portugal}
\affiliation{Departamento de F\'{\i}sica, Faculdade de Ci\^{e}ncias,
Universidade de Lisboa, 1749-016 Lisboa, Portugal}  

\begin{abstract}
We study the spontaneous folding of a 2D template of microscopic panels into a
3D pyramid, driven by thermal fluctuations. Combining numerical simulations
and analytical calculations, we find that the total folding time is a
non-monotonic function of the number of faces, with a minimum for five faces.
The motion of each face is consistent with a Brownian process and folding
occurs through a sequence of irreversible binding events that close edges
between pairs of faces. The first edge closing is well-described by a
first-passage process in 2D, with a characteristic time that decays with the
number of faces. By contrast, the subsequent edge closings are all
first-passage processes in 1D and so the time of the last one grows
logarithmically with the number of faces. It is the interplay between these
two different sets of events that explains the non-monotonic behavior.
Possible implications in the self-folding of more complex structures are
discussed. 
\end{abstract}

\maketitle

Kirigami is the art of cutting two-dimensional templates and fold them into
three-dimensional structures. Nowadays, there is a growing interest on
extending this ancient idea to design materials that fold spontaneously into
targeted 3D structures. The driving mechanism depends on the lengthscale. At
the macroscale, folding is driven by energy minimization (e.g. stress
relaxation), and thus the folding pathway is
deterministic~\cite{shenoy2012self,liu2012self,erb2013self,castle2014making,sussman2015algorithmic,dudte2016programming,liu20162d,liu2017sequential,paulsen2019wrapping,dieleman2019jigsaw,santangelo2019fold,siefert2019bio}.
By contrast, at the microscale, since folding occurs usually in suspension,
the fluctuations in the fluid-structure interaction dominate and folding is
stochastic~\cite{pandey2011algorithmic,dodd2018universal}. This challenges the
use of Kirigami at the microscale as, for example, in encapsulation, drug
delivery, and soft
robotics~\cite{fernandes2012self,shim2012buckling,filippousi2013polyhedral,felton2014method}.

To design self-folding Kirigami, one first needs to identify what are the
two-dimensional templates (nets) that fold into the desired structure. For
shell-like structures of rigid panels connected by edges, these nets are
obtained by edge unfolding, i.e., by cutting edges and opening the
structure~\cite{demaine2007geometric}. In principle, different nets can fold
into the same three-dimensional structure. However, recent experiments and
numerical simulations show that the stochastic nature of folding might lead to
misfolding. By performing independent samples, they found that the probability
for a given net to fold into the desired structure (yield) strongly depends on
the topology of the net and experimental
conditions~\cite{azam2009compactness,pandey2011algorithmic,araujo2018finding,dodd2018universal}.
Thus, the focus has been on identifying what are the optimal nets that
maximize the yield~\cite{pandey2011algorithmic,araujo2018finding}. But, what
about the folding time? For practical applications, it is not only critical to
reduce misfolding but also to guarantee that folding occurs in due time. Here,
we address this question. To focus on the folding time, we consider as a
prototype the spontaneous folding of a pyramid, where misfolding is not
possible.
\begin{figure}
\includegraphics[width=1.0\columnwidth]{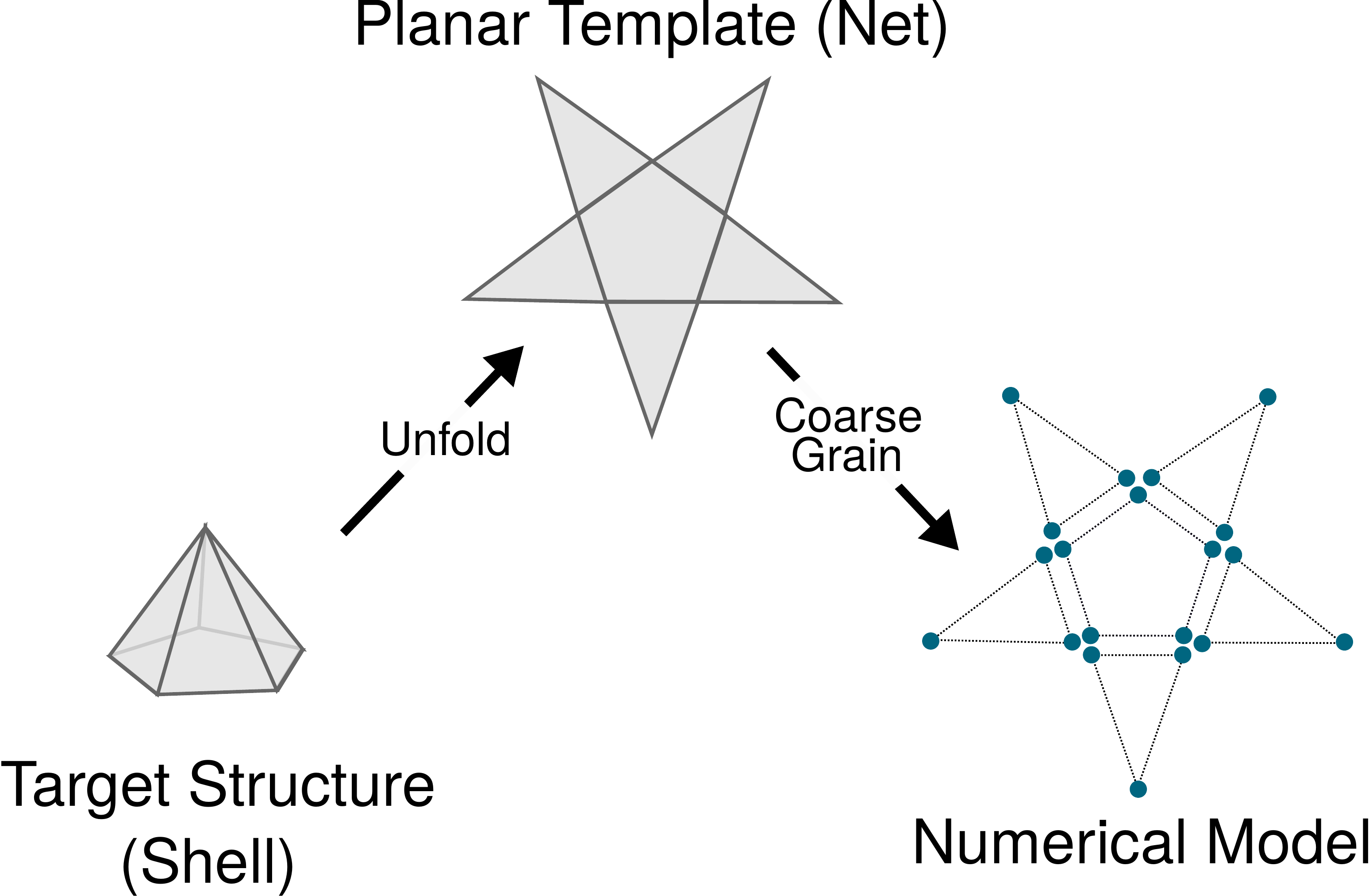}
\caption{\textbf{Unfolding process and model.} We consider pyramids (left) of
one base and $N$ lateral faces ($N=5$ in the figure). The 2D template of
microscopic panels (center) is obtained by cutting the edges between the
lateral faces and unfolding the faces. To simulate the folding dynamics, we
developed a coarse-grained numerical model where each face is described as a
rigid body of three particles (right) at the vertices. The base is described
by $N$ particles at the vertices.  The interaction between particles is
considered pairwise and attractive. To suppress misfolding, the base is pinned
to a flat substrate and the lateral faces can only fold in one
side.~\label{fig.model}}
\end{figure}

Let us consider a pyramid with $N$ lateral faces (see Fig.~\ref{fig.model}).
The 2D net is a $N$-pointed star, obtained by cutting the edges of the lateral
faces and unfolding them. To simulate the folding dynamics, as explained in
detail in the Supplemental Material~\cite{SM} and summarized in
Fig.~\ref{fig.model}, we performed particle-based simulations. We are
interested in the limit where the interaction between faces is short-ranged
(contact like) and the edge closing irreversible. Thus, each face is described
as a rigid body of three particles at the vertices. The attractive interaction along the edges
is modeled by a strong inverted-Gaussian potential between particles.
The stochastic trajectories of the faces under thermal fluctuations
is obtained by solving the corresponding Langevin equations, where the noise
term is parameterize by a rotational diffusion coefficient $D_0$ of the
lateral faces. To suppress misfolding, we pinned the base of the pyramid to a
substrate and so the faces can only fold in one side (see Supplemental
Material~\cite{SM} for further details). 

\begin{figure}
\includegraphics[width=1.0\columnwidth]{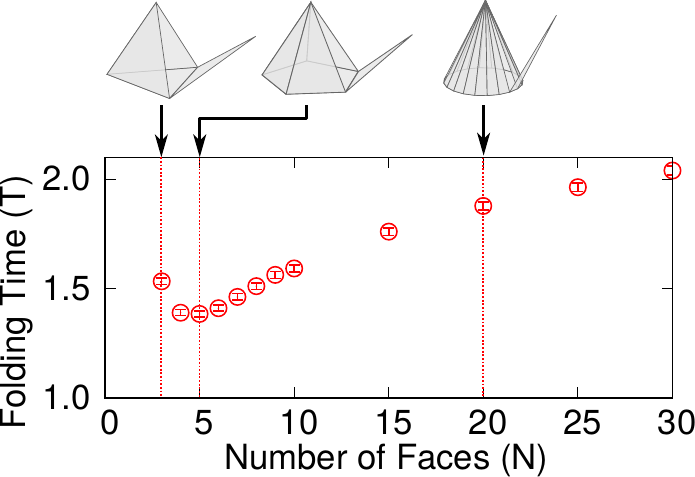}
\caption{\textbf{Non-monotonic dependence of the folding time on the number of
lateral faces.} Folding time as a function of the number of lateral faces
($N$), defined as the total time necessary for all faces to fold into a
pyramid. Time is in units of Brownian time, i.e., the average time for a
non-interacting face to diffuse over an angular region of size $\pi$. Results
are obtained numerically by averaging over $2\times10^3$ independent samples,
starting from a flat template.~\label{fig.FoldingTime}}
\end{figure}
\begin{figure*}[t]
\includegraphics[width=1.8\columnwidth]{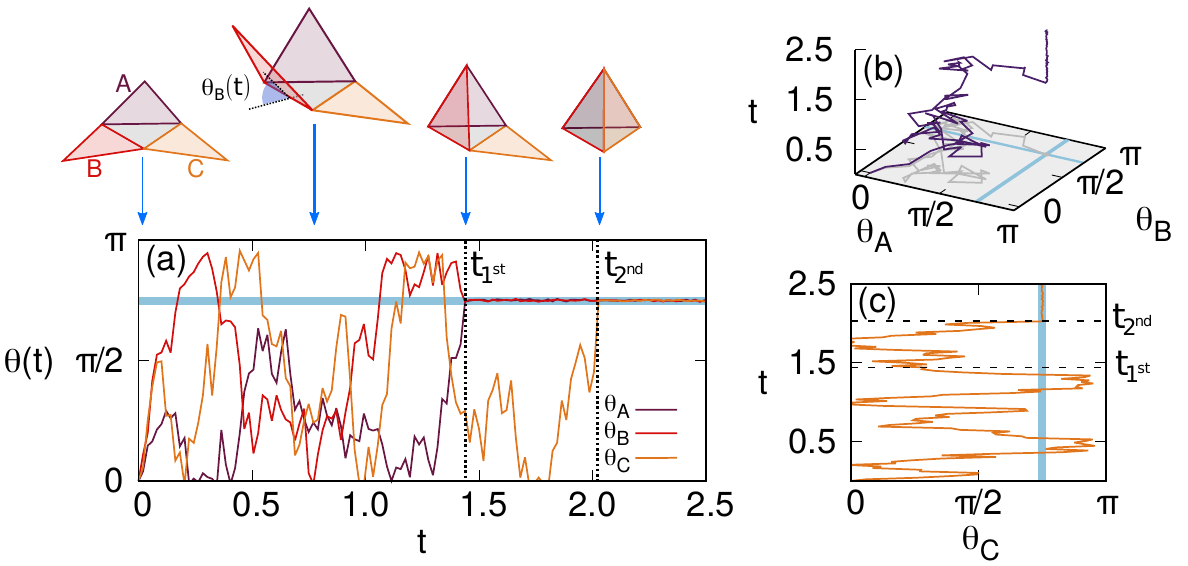}
\caption{\textbf{Self-folding of a pyramid of $N=3$ lateral faces.} (a) Time
dependence of the angle $\theta_i$ of each face $i$ with the flat substrate (see
scheme). Since the base of the pyramid is pinned and the faces can only move
in one side, $\theta_i\in\left[0,\pi\right]$. Results are for one sample,
starting from a flat template ($\theta_i(0)=0$). The statistics of $\theta_i$
is consistent with a Brownian process with reflective walls at $\theta_i=0$
and $\theta_i=\pi$ (see Supplemental Material~\cite{SM}) and folding occurs
through a sequence of edge closings. For the first edge closing, two
faces need to be at $\theta^*=3\pi/4\pm\Delta$ (region in blue) at the same
time, where $\Delta\approx\pi/180$. In the example, this occurs for faces $A$
and $B$ at time $t_{1^\mathrm{st}}$. (b) If we assume no interaction between
$A$ and $B$ outside the blue region, the first edge closing can be mapped into
a 2D Brownian process, with coordinates $\theta_A$ and $\theta_B$ and a trap
in a region where both $\theta_i=\theta^*$. $t_{1^\mathrm{st}}$ is then the
first-passage time. (c) After the first edge closing, the third face $C$,
performs a 1D Brownian motion until it hits $\theta^*$. In the example, this
occurs at $t_{2^\mathrm{nd}}$. As in Fig.~\ref{fig.FoldingTime}, time is
rescaled by Brownian time.~\label{fig.Browniano}}.
\end{figure*}

We performed independent simulations for different numbers of lateral faces
$N$, starting from a flat (2D) configuration and running until the final
pyramid is obtained. As shown in Fig.~\ref{fig.FoldingTime}, we find that, the
total folding time $T$ is a non-monotonic function of the number of faces $N$,
with an optimal time for five faces. To characterize the dynamics, we define
$\theta_i$ as the angle between the face $i$ and the substrate (see scheme in
the top of Fig.~\ref{fig.Browniano}). Since the motion is constrained by the
substrate, $\theta_i\in\left[0,\pi\right]$. 

As an example, we consider now the folding of a pyramid of three lateral faces
($N=3$). The time dependence of the three angles is shown in
Fig.~\ref{fig.Browniano}(a).  Due to thermal fluctuation, each face jiggles
around until the first two faces ($A$ and $B$ in the figure) meet at time
$t_{1^\mathrm{st}}$ and bind irreversibly, closing the edge between them. The
third face ($C$) also binds to the first two at a later time
$t_{2^\mathrm{nd}}$. Thus, folding occurs through a sequence of irreversible
edge closings. Below, we discuss the first and subsequent edge closings
independently.

\begin{figure}
\includegraphics[width=1.0\columnwidth]{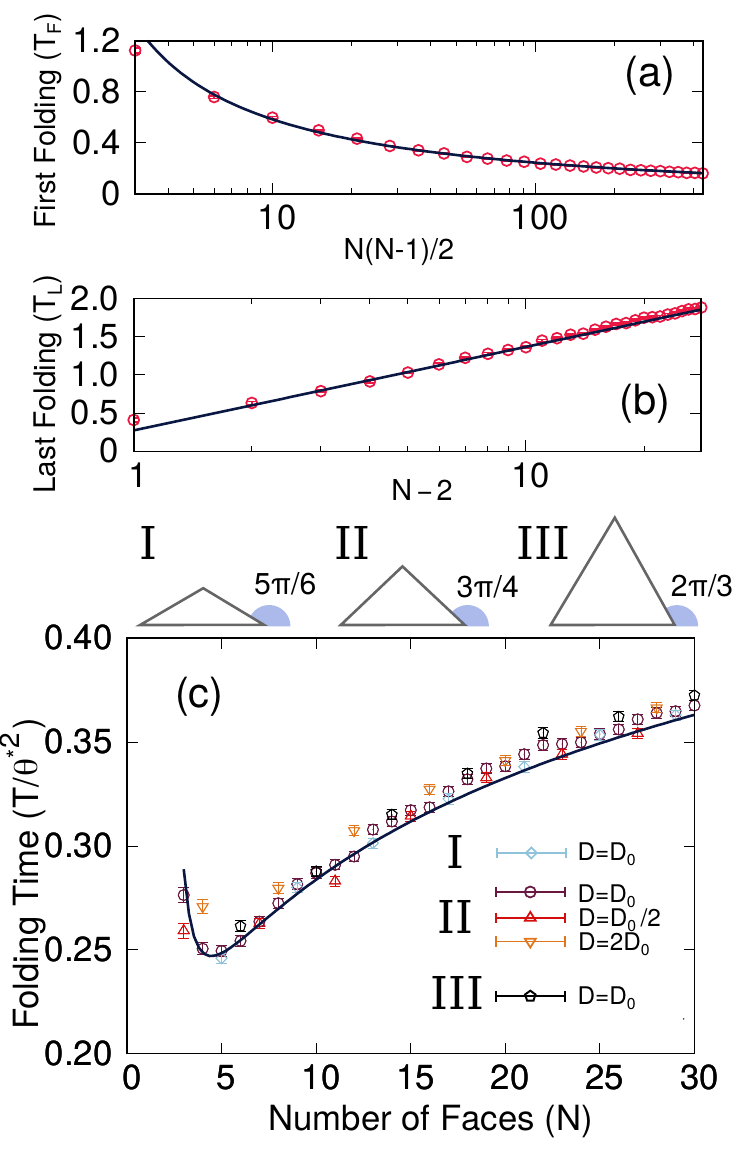}
\caption{\textbf{Dependence on the number of lateral faces.} (a) Time of first
edge closing ($T_F$) as a function of $N(N-2)/2$, where $N$ is the number of
lateral faces. The first edge closing occurs when two faces have
$\theta=\theta^*=3\pi/4\pm\Delta$, where $\Delta\approx\pi/180$. $T_F$
decreases with $N$. The solid line is given by $T_{F} =
\tau_F/\ln(N(N-1)/2)+\tau_{F0}$, where $\tau_F=1.57\pm 0.02$ and
$\tau_{F_0}=-0.097\pm 0.006$ are fitting parameters obtained by the least
square fit of the simulation data. This expression corresponds to a
first-passage time of a 2D Brownian process (see details in the text). (b)
Last-edge-closing time as a function of $N-2$. After the first edge closing,
$N-2$ faces remain that fold sequentially. $T_L$ is given by the slowest of
($N-2$) 1D first-passage events. The solid line is given by $T_{L} = \tau_L
\ln(N-2)+ \gamma \tau_L$, with $\tau_L=4{\theta^*}^2/D_0\pi^2$ and $\gamma$
the Euler-Mascheroni constant.  $D_0$ is the angular diffusion coefficient of
reference (see Supplemental Material~\cite{SM}).  (c) Data collapse for the
folding time, in units of Brownian time, rescaled by the square of the folding
angle $(\theta^*)^2$, as a function of the number of faces $N$ for different
values of $\theta^*=\{3\pi/4, 2\pi/3, 5\pi/6\}$ and $D=\{D_0, D_0/2,
2D_0\}$.  Results are averages over $2\times10^3$ independent
samples.~\label{fig.FirstLast}}
\end{figure}
As shown in the Supplemental Material~\cite{SM}, the statistics of the three
time series $\theta_i(t)$ in Fig.~\ref{fig.Browniano} is consistent with a 1D
Brownian process with reflective boundaries at $\theta_i=0$ and
$\theta_i=\pi$. The short-ranged (attractive) interaction between faces is
only effective in a small region of the angular space,
$\theta^*=3\pi/4\pm\Delta$, with $\Delta\approx\pi/180$ as estimated from the
properties of the potential (see Supplemental Material~\cite{SM}). For the
first edge closing to occur, the angle of two faces need to be at $\theta^*$
at the same time and, once there, they get trapped. Thus, if we map the motion
of each pair of faces $j$ and $k$ into a 2D Brownian process, with coordinates
$(\theta_j,\theta_k)$ and a trap at $(\theta^*,\theta^*)$, the edge closing
between $j$ and $k$ occurs when the corresponding 2D Brownian process hits the
trap (see Fig.~\ref{fig.Browniano}(b)). In the general case of $N$ lateral
faces, since there are $N(N-1)/2$ possible pairs of faces, the time of the
first edge closing is the fastest of $N(N-1)/2$ first-passage processes.

To estimate the average time $T_F$ of the first edge closing for a pyramid of
$N$ lateral faces, we define $g(t)$ as the first-passage time distribution of
a 2D Brownian process. There are $N(N-1)/2$ pairs of faces and so the same
number of competing Brownian processes. The first edge closing is the fastest
of all possible ones and thus $T_F=min\{t_1,t_2,t_3...t_{N(N-1)/2}\}$, where
$t_i$ are random values following the distribution $g(t)$. If we neglect any
correlations between the motion of the different faces, from the theory of
order statistics~\cite{arnold1992first}, we estimate that,
\begin{equation}\label{eq.first_meeting}
T_F(N) = \frac{N(N-1)}{2}\int_0^\infty{t g(t) \left[\int_t^\infty
g(t')dt'\right]^{\frac{N(N-1)}{2}-1} dt} \ \ ,
\end{equation} 
where the term with the square brackets corresponds to the probability that,
provided that a first-passage process occurs at time $t$, all the remaining
$N(N-1)/2-1$ occur at a later time. $g(t)$ depends on the geometry and initial
conditions~\cite{yuste1996order,weiss1983order,holcman2014narrow,singer2006narrow,yuste1997escape}.
For a set of $N(N-1)/2$ Brownian
processes~\cite{weiss1983order,basnayake2019asymptotic} starting at the origin
($\theta_i(0)=0$), 
\begin{equation}\label{eq.first_meeting_here}
T_F(N)\sim 1/\ln\left(\frac{N(N-1)}{2}\right) .
\end{equation}
So, the time of the first edge closing should decrease with the number of
possible pairs. Figure~\ref{fig.FirstLast}(a) shows $T_F$ in units of Brownian
time (see figure caption), for different numbers of lateral faces, obtained
numerically by averaging over $10^4$ samples. The solid line is given by
$T_F=\tau_F/\ln\left(N(N-1)/2\right)+\tau_{F_0}$, where $\tau_F=1.57\pm 0.02$ and
$\tau_{F_0}=-0.097\pm 0.006$ are obtained by fitting the simulation data.
Clearly, the decrease in $T_F$ with the number of faces is well described by
Eq.~\eqref{eq.first_meeting_here}.

The dynamics of the subsequent edge closings is fundamentally different. While
for the first edge closing, two faces need to meet at a particular angular
$\theta^*$, the remaining faces will close edges one-by-one as soon as they
reach $\theta^*$. The folding is complete when all faces reach this value.
Thus, each of the subsequent ($N-2$) edge closings is a 1D first-passage
process (see Fig.~\ref{fig.Browniano}(c)). We define $T$ as the total folding
time and $T_L=T-T_F$ as the time from the first to the last edge closing. Each
free face $i$ binds when $\theta_i(T_F+t)=\theta^*$ (with $t\geq0$) for the
first time. To estimate $T_L$, we assume that $\theta_i(T_F)<\theta^*$ for all
$i$ and that $\theta_i(T_F+t)$ is well described by a 1D Brownian process,
with one reflective boundary at $\theta_i=0$ and a trap at $\theta^*$. $T_L$
is then the slowest of the ($N-2$) 1D first-passage processes. Thus,
\begin{equation}
T_L(N) = (N-2)\int_0^\infty{t f(t) \left[1-\int_t^\infty
f(t')dt'\right]^{(N-2)-1} dt} , \label{eq.last_meeting}
\end{equation}
where $f(t)$ is the 1D first-passage time distribution and the term with
square brackets is the probability that, provided that a first-passage process
occurs at time $t$, all the remaining ones were faster. Assuming that
$\theta_i(T_F)$ is uniformly distributed in $\left[0,\theta^*\right]$, the
first-passage time distribution is $f(t)\approx e^{-t/\tau_L}$, with $\tau_L=4
{\theta^*}^2/D_0\pi^2$~\cite{redner2001guide}, where $D_0$ is the diffusion
coefficient of the Brownian process. This gives,
\begin{equation}
T_L(N) = \tau_L \sum_{i=1}^{N-2}\frac{1}{i} , \label{eq.last_meeting_2}
\end{equation}
and thus, $T_L(N)\approx\tau_L\left[\ln(N-2) + \gamma\right]$, where $\gamma$
is the Euler-Mascheroni constant. Figure~\ref{fig.FirstLast}(b) depicts $T_L$
obtained numerically for different $N$. The numerical data is consistent with
Eq.~\eqref{eq.last_meeting_2} (solid line).

The dependence on the number of lateral faces $N$ of $T_F$ and $T_L$ is
significantly different. While $T_F$ decreases with $N$, $T_L$ grows. The
total folding time $T$ is the sum of the two. Thus, for low values of $N$, the
total folding time is dominated by the time of the first edge closing, whereas
for large $N$ is the last closing that sets the overall timescale. It is the
interplay between these two timescales that explains the minimum observed in
Fig.~\ref{fig.FoldingTime}. 

So far, we considered always the same closing angle $\theta^*$ and diffusion
coefficient $D_0$. Since the motion of the faces is diffusive, all timescales
should scale with $\tau={\theta^*}^2/2D_0$, which is the average time for a
non-interacting 1D Brownian process to diffuse in an angular region of size
$\theta^*$. Figure~\ref{fig.FirstLast}(c) shows the folding time obtained
numerically for different values of $\theta^*=\{3\pi/4, 2\pi/3, 5\pi/6\}$ and
$D=\{D_0, D_0/2, 2D_0\}$. A data collapse is obtained when time, initially
in units of Brownian time is rescaled by $(\theta^*)^2$. The solid line is the
sum of the solid lines for $T_F$ and $T_L$ in Figs.~\ref{fig.FirstLast}(a) and
(b), respectively, and describes quantitatively the dependence on the number
of lateral faces.

\emph{Conclusions.} Under thermal fluctuation, a $N$-pointed star template of
rigid panels and flexible hinges folds into a 3D pyramid of $N$ lateral faces.
Folding occurs through a sequence of edge closings, but the nature of the
first and subsequent edge closings is significantly different. For the first
edge closing, two jiggling faces need to meet at a particular angle, whereas
for the subsequent edge closings, only one face needs to reach that angle. We
hypothesized that the first edge closing can be mapped into a first-passage
event of a 2D Brownian
process~\cite{holcman2015stochastic,schuss2007narrow,singer2006narrow,holcman2014narrow},
obtaining an expression for the corresponding time. This expression predicts
that the time for the first edge closing decreases with $N$, what describes
quantitatively the numerical data. By contrast, to estimate the time for the
subsequent edge closing, we mapped them into a set of first-passage events in
1D and derived the time for the slowest of them all. We predict that this time
should rather grow logarithmically with $(N-2)$, which is also observed
numerically. Since the total folding time is the sum of the two times, a
non-monotonic dependence on $N$ is found.

Spontaneous folding at the microscale is an intricate process that might
depend on the physical properties of the structure, fluid-structure
interactions, and thermostat
temperature~\cite{pandey2011algorithmic,dodd2018universal,araujo2018finding}.
Nevertheless, our approach shows that, by mapping folding into a set of
competing Brownian processes and binding events, one can predict accurately
the relevant time scales. For simplicity, we considered a pyramid, a structure
with equivalent folding panels. In general, the template for a given
polyhedral structures has different types of panels. They differ not only in
shape and size, but also in their position relative to the panel of reference
(e.g.  base). To extend our framework to those structures, it is critical to
consider that folding evolve through a hierarchy of edge closing events that
depend on the kinetic pathway of folding. 

\emph{Acknowledgments.} We acknowledge financial support from the Portuguese
Foundation for Science and Technology (FCT) under Contracts nos.
PTDC/FIS-MAC/28146/2017 (LISBOA-01-0145-FEDER-028146), UIDB/00618/2020,
UIDP/00618/2020, and CEECIND/00586/2017.

\bibliography{paper}

\end{document}